\documentclass[runningheads]{llncs}
\usepackage[T1]{fontenc}
\usepackage{graphicx}
\usepackage{subfigure}

\title{From Documents to Database: Failure Modes for Industrial Assets}

\author{Duygu Kabakci-Zorlu\orcidID{0000-0001-6636-813X} \and
Fabio Lorenzi\orcidID{0009-0006-1113-2377} \and
John Sheehan \and
Karol Lynch
\and Bradley Eck}

\authorrunning{Kabakci-Zorlu et al.}
\institute{IBM Research Europe, Dublin, Ireland \\ 
\email{duygu.kabakci.zorlu@ibm.com, fabio.lorenzi1@ibm.com, john.d.sheehan@ie.ibm.com, \\ karol\_lynch@ie.ibm.com,   bradley.eck@ie.ibm.com} }

\begin{document}

\maketitle

\begin{abstract}
We propose an interactive system using foundation models and user-provided technical documents to generate Failure Mode and Effects Analyses (FMEA) for industrial equipment.
Our system aggregates unstructured content across documents to generate an FMEA and stores it in a relational database.
Leveraging this tool, the time required for creation of
this knowledge-intensive content is reduced, outperforming traditional manual approaches.  
This demonstration showcases the potential of foundation models to facilitate the creation of specialized structured content for enterprise asset management systems.
\keywords{Foundation Models \and Structured Text Generation \and FMEA \and Reliability.}
\end{abstract}

\section{Introduction}

Failure Mode and Effects Analysis (FMEA) is a well-established 
technique in reliability engineering for identifying an equipment's potential failure points and determining optimal maintenance strategies~\cite{srt2004,sharma:JARASS-2018}. 
These documents, which record the reasons for equipment failures and outline maintenance options, play a crucial role in managing physical assets to 
a desired level of reliability.  

FMEAs are extensive documents containing information
from multiple sources including text, diagrams, tables
and--crucially--the knowledge and experience of reliability professionals. 
The usual approach of FMEA creation consists of workshops facilitated among experts to gather and discuss the material. Resources needed for this approach can amount to millions of dollars across an enterprise~\cite{cooper2015FMEAs}.

Compiling data from multiple sources into a structured database is a task that requires significant time and effort. The structured representation of FMEA allows consistency across equipment types and enables integration with asset management systems. 
Our approach targets tree structured FMEAs of depth five for each study. First,
the \emph{boundary} provides a 
functional overview and identifies the main parts of the system. Second, the \emph{failure locations} pinpoint locations where potential failures might happen.
Third, physical processes contributing to failures are termed
\emph{degradation mechanisms}. Fourth, \emph{degradation influences} note the underlying causes of the degradation, known as degradation influences. 
A triple of failure location, degradation mechanism and degradation influence is termed \emph{failure mode}. Fifth, to
anticipate and avert these potential issues, the document provides \emph{preventive maintenance tasks}. 

This hierarchical structure poses several challenges. The sequential relationship between the sections propagates errors: an incorrect failure location will also have incorrect degradation mechanisms. 
Expert guidance between steps is crucial to prevent cascading errors.
Additionally, FMEAs contain domain-specific knowledge about the equipment and its usage, making it difficult to create a catalog of components for building FMEAs due to the ambiguity of certain terms e.g., the term ``casing'' refers to different components in a pump and in a window. 
Consolidating knowledge from domain experts and documents into this representation is the scope of this work.

Since FMEA generation relies on considerable labor of domain experts, earlier workers have also sought to bring the productivity improvements of AI to the problem.
~\cite{Hodkiewicz2021} proposes an ontology-based approach to automate reasoning on  FMEA spreadsheets. 
\cite{fmea_builder_ijcai_2024} and ~\cite{Lynch2025} propose an LLM based system for creating new FMEAs with in-context learning using an expert crafted database and novel ensemble strategies. Another Retrieval Augmented Generation \cite{lewis2021rag}
(RAG) based multi-agent LLM application leverages a structured database as external data source \cite{Xia2024}. Additionally, a FMEA generation framework with a fine-tuned LLM from domain specific data is proposed by ~\cite{elHassani2024}.
So far, AI powered creation of FMEAs from source documents remains unexplored and is the focus of this paper.

In this work, we consider the problem of interactive FMEA generation from 
unstructured documents. The problem is relevant where the asset of interest differs substantially from available knowledge data as used in other approaches.
We show an expert supervised and interactive LLM based system for accelerating structured FMEA creation from source documents.
To the best of our knowledge, this system contributes a first example 
of structured information extraction from documents in this domain.

\section{System overview}
\begin{figure}
\includegraphics[width=0.9\textwidth]{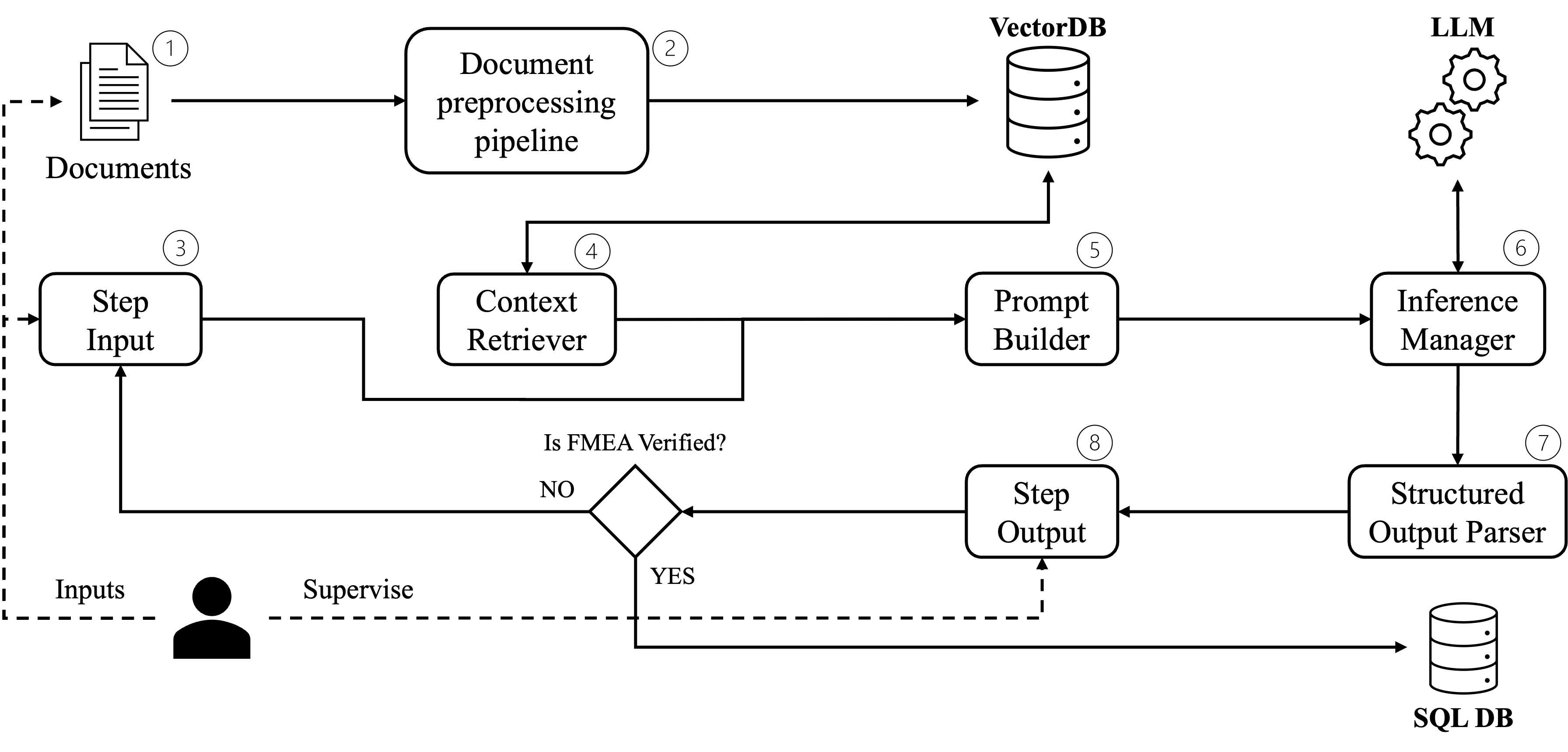}
\caption{System flow diagram}
\label{fig:flow}
\end{figure}

The system builds upon the work from \cite{fmea_builder_ijcai_2024} \cite{Lynch2025} that offers structured text generation with in-context learning from an expert crafted FMEA library. We recognize that integration of unstructured documents can expedite the generation process in cases where the targeted asset differs substantially from available existing FMEAs.  We follow a RAG approach to ground LLM responses with external context.
External context comes from user provided documents such as operation and maintenance manuals for the asset. These documents typically include information about installation, operation, inspection, and maintenance of an equipment, therefore containing considerable amount of relevant information.
The system flow (Figure~\ref{fig:flow}) is divided into two primary phases: 1. LLM-assisted document pre-processing; and, 2. Structured document generation with user supervision.

\label{sec:approach}

\subsection{Document Preprocessing}
\begin{figure}
\includegraphics[width=0.8\textwidth]{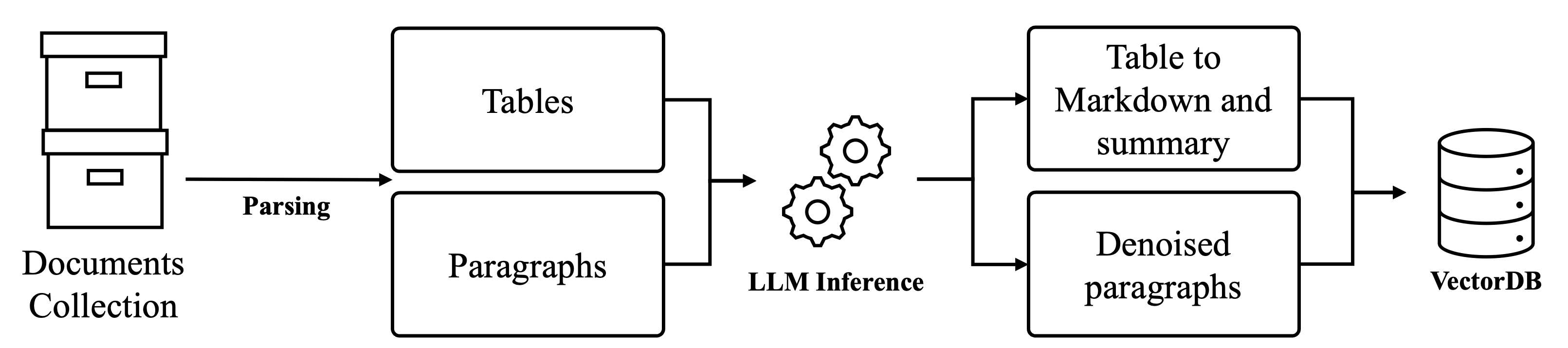}
\caption{LLM-assisted document pre-processing diagram}
\label{fig:preprocess-flow}
\end{figure}

The LLM-assisted document pre-processing pipeline (Figure \ref{fig:preprocess-flow}) extracts information from unstructured documents such as PDFs using the Docling library ~\cite{auer2024doclingtechnicalreport}. 

\label{sec:preprocess-approach}

\begin{figure*}
    \centering
\subfigure{\includegraphics[width=0.27\textwidth]{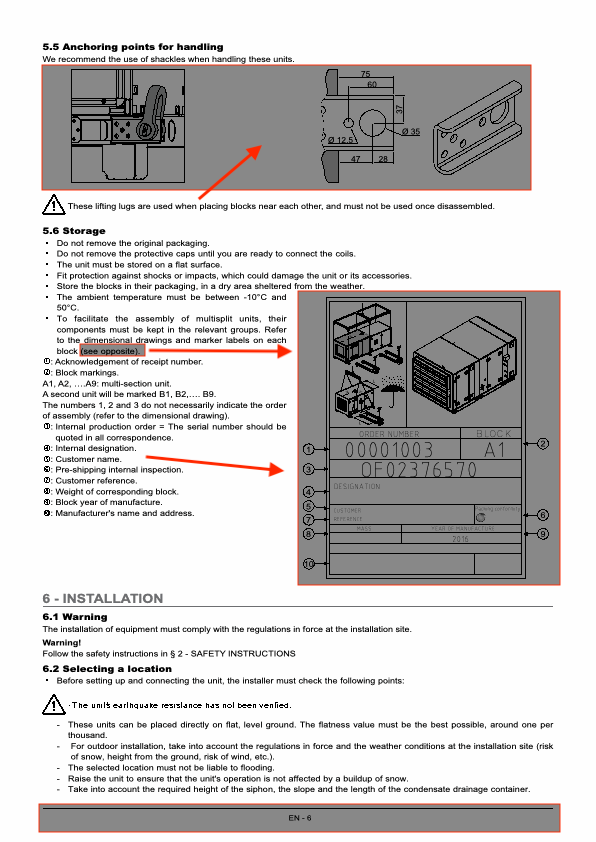}}
    \subfigure{\includegraphics[width=0.27\textwidth]{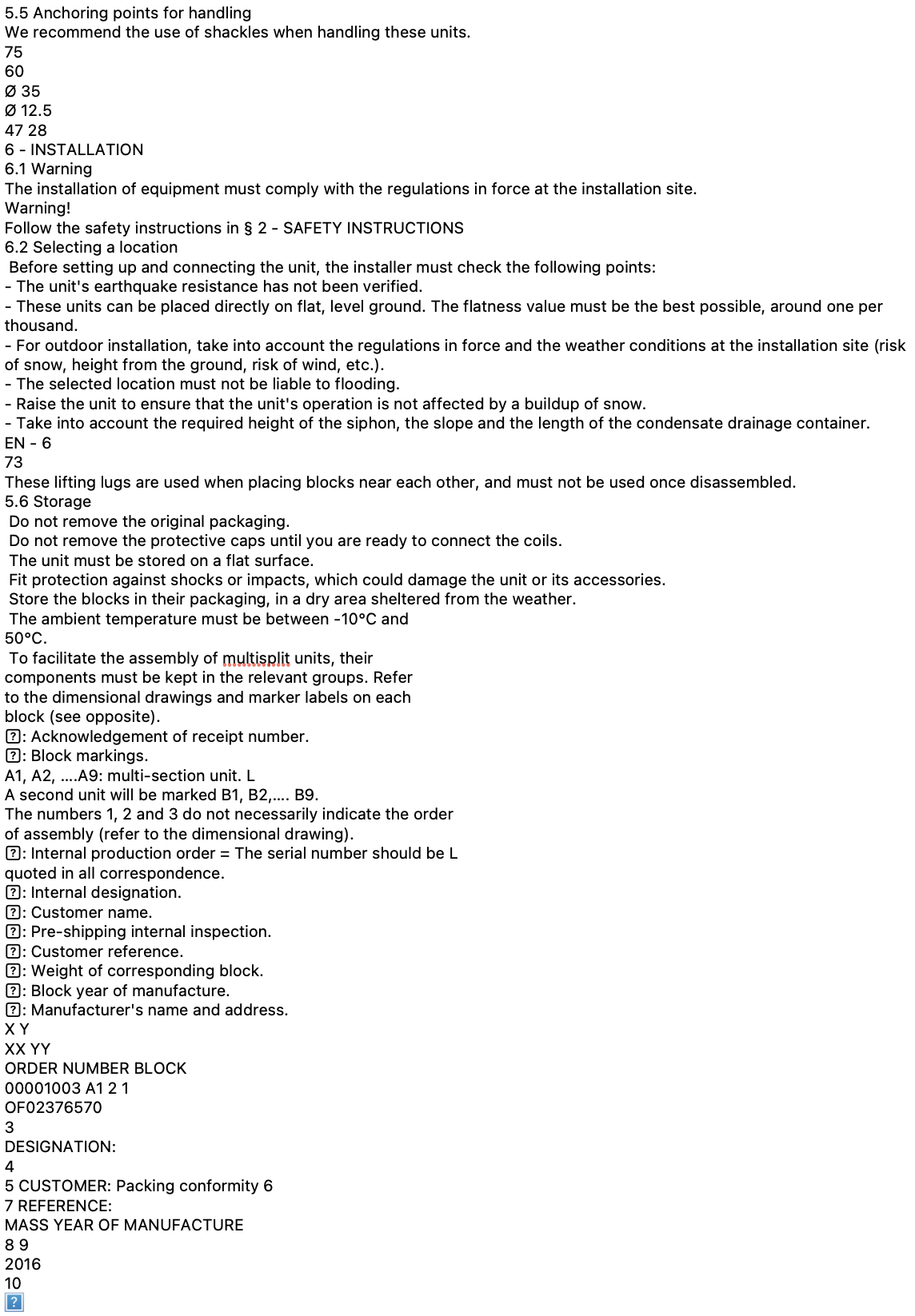}} 
 
    \subfigure{
        \includegraphics[width=0.3\textwidth]{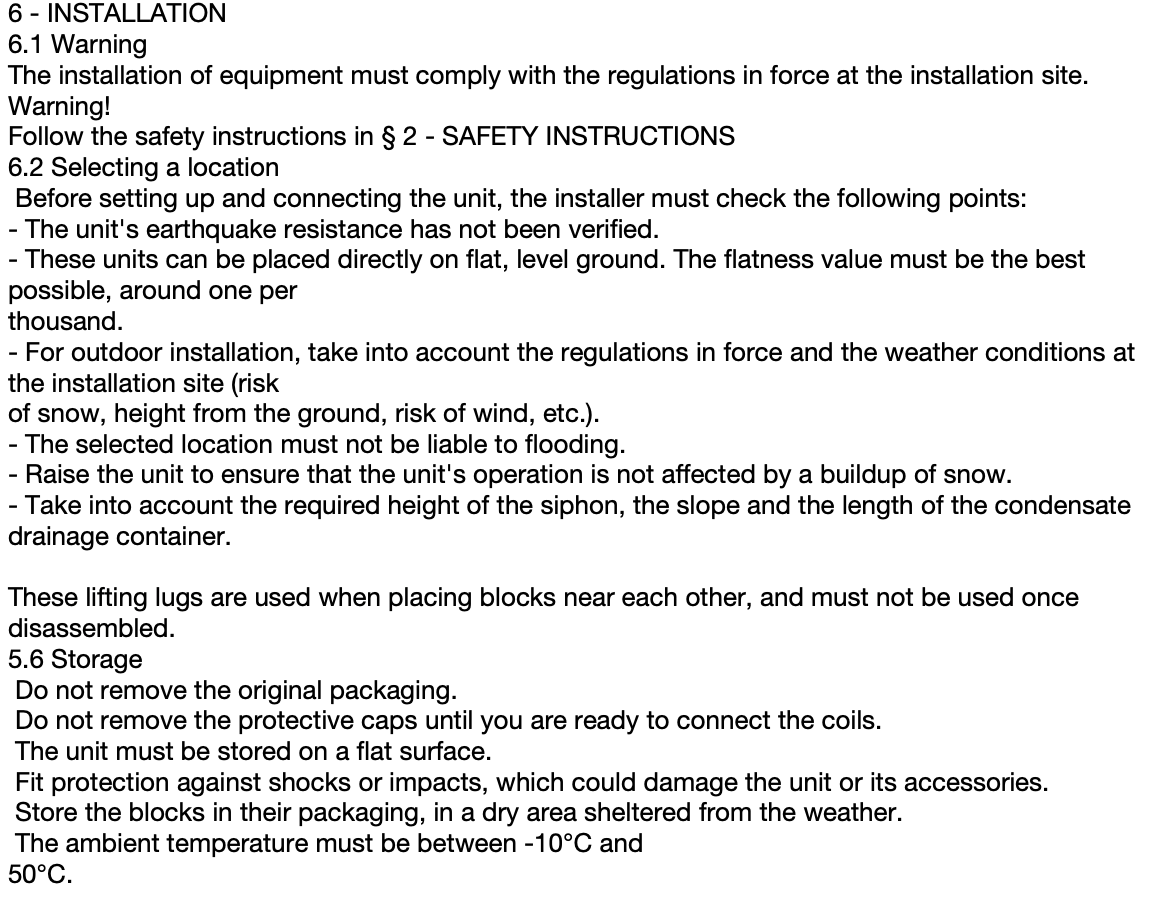}}
    \caption{ Example of document page with noisy and  cleaned chunks before and after filtering}
    \label{fig:noise_example}
\end{figure*}

Document processing, for PDFs in particular, presents some challenges in the form of noisy text extraction and semantic interpretation of tables. Noise in this particular case is identified as any textual information which is not part of the document corpus or that can't be interpreted without non-textual references e.g., headers and footers, footnotes, captions, page numbers. (Figure~\ref{fig:noise_example}). 

The noise reduction pipeline uses an LLM specifically prompted to filter the raw text into cohesive paragraphs. The resulting paragraphs are chunked and indexed in a vector database. Tables are parsed out of documents alongside captions, if any, and converted to markdown format before being sent for summarisation by an LLM. The tables are stored in a vector database indexed by their summary. The content of the database after the processing pipeline provides contextual knowledge during the FMEA generation rounds.

\subsection{System flow for supervised generation}

Our system follows the flow of 
Figure~\ref{fig:flow} to generate FMEA documents with user supervision. The process goes through nine steps.
The user uploads one or many documents containing domain knowledge about the asset (1); these documents are pre-processed (2) and indexed in vector DB as described in Figure~\ref{fig:preprocess-flow}. The user initiates the FMEA generation by providing some asset information such as a short description of the asset (3). The user input is encoded as a query to the vector database; the top-k matching text chunks from the processed documents are returned (4) and compiled into engineered prompts containing generation instructions and hints about the desired output structure (5). 

The prompts are handed over to an inference manager which interacts with an instance of an LLM. FMEA documents are presented in a tree structure.  An equipment type, its components and their degradation mechanisms are in one to many relationships. To mirror this structure, our application executes sequential generative steps. To ensure required structured outputs, the generated text is passed through a rule-based parser that cleans the response and extracts a structured object from it (7); the structured response is presented to the user for further editing and supervision (e.g., adding list items or modifing part of generated text) (8). The structured nature of the responses also facilitates insertion of the final FMEA content into a relational database (9), allowing for integration with other FMEA related tools.

\begin{figure*}
    \centering
    \subfigure{\includegraphics[width=0.45\textwidth]{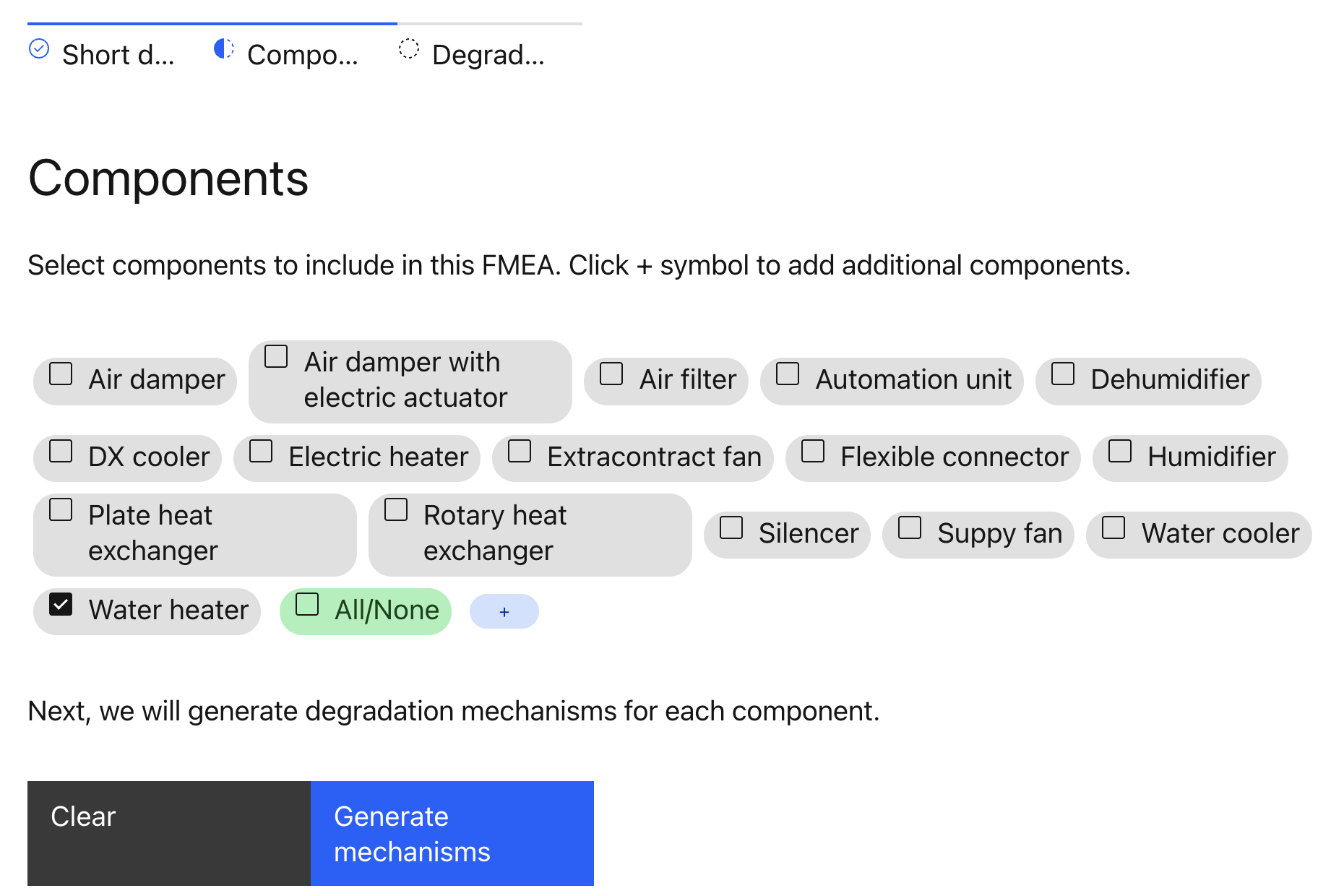}} 
    \subfigure{
        \includegraphics[width=0.4\textwidth]{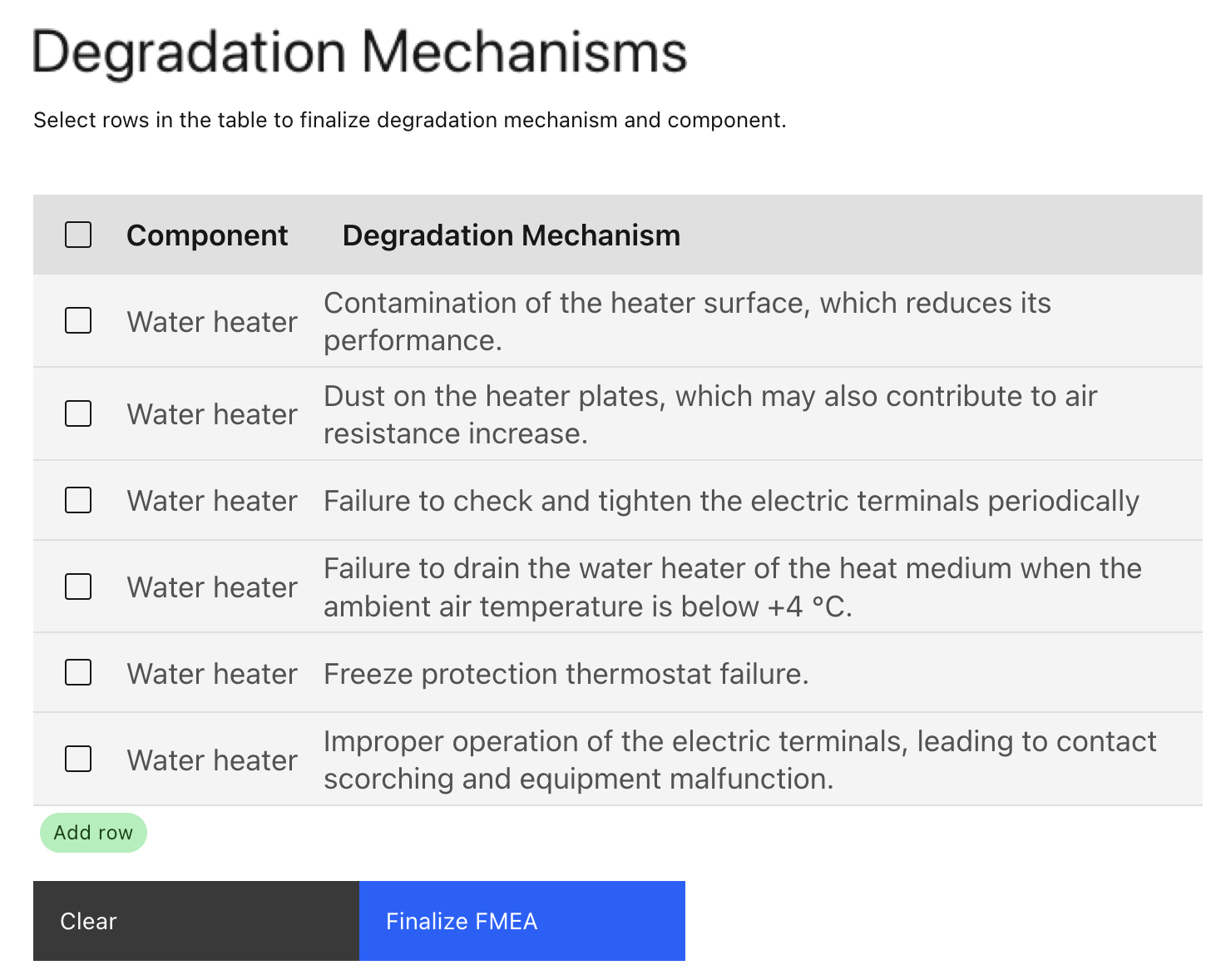}}
    \caption{ UI Visuals for Air Handling Unit (a) Components step  (b) Degradation Mechanisms step}
    \label{fig:fmea-ui-visuals}
\end{figure*}

Interaction with the system happens through a user interface (UI) that allows users to upload and store documents. 
In the example of Figure \ref{fig:fmea-ui-visuals}a, the
user selects relevant documents and generates a list of maintainable components for an air handling unit. Next, the user generates the associated degradation mechanisms for the selected components (Figure \ref{fig:fmea-ui-visuals}b).

\section{Experiments}

Our evaluation focuses on the \emph{failure locations} extracted from guide documents
of 12 equipment types. We compare our proposed method against the zero-shot prompting approach. We also analyze retrieval quality from
document chunks and the effect of the context length. 

\subsection{Data Preparation}
\label{sec:data_prep}
We synthesized the guide documents to mimic real asset documents containing a mix of relevant and irrelevant information in both body text and tables. The guides contained the failure locations of interest along with other parts of the FMEA such as degradation mechanisms and influences in the main text. We also added sections of LLM generated content related to installation and troubleshooting. Less relevant information included 
tables of failure locations and failure modes for similar equipment than the one being evaluated. 
The 12 guide documents analyzed here had an average of 26 pages and covered miscellaneous asset types e.g., hammer drill. On average, each asset had 21 failure locations.

\subsection{Performance Evaluation}

Performance evaluation focused on identifying failure locations. This crucial part of the FMEA records the equipment's structure thus influences subsequent fields.
These experiments  
compare  failure locations returned from guide documents for 12 equipment types. 
Golden failure locations came from our database of expert crafted FMEAs.  
We considered four scenarios:  (i) zero-shot, using no documents as an experimental baseline; 
(ii) 3 chunks, where the top three relevant chunks from the document are included in the prompt; (iii) 5 chunks, 
and (iv) long context, where the entire guide document appears in the prompt.

Experiments used \emph{msmarco-MiniLM-L-6-v3}\footnote{http://huggingface.co/sentence-transformers} as an embedding model and Granite-3-8B-Instruct \cite{granite2024ver3} (6) for LLM inferencing.
Chunks had maximum length of 1024 characters. 
Reported metrics include Structured Semantic Entity Evaluation (SSEE) precision and recall as introduced in ~\cite{Lynch2025}. 
These metrics measure the fraction of correctly identified entities from candidate and gold set by comparing generated list of failure locations and gold set in a shared embedding space with similarity threshold.

Results (Table \ref{tab:fl_results}) showed that all variations of our system strongly outperformed the zero-shot baseline.
Zero-shot performance is low, but not zero, indicating that LLMs do contain some "specialised" knowledge. However, providing reference documents with relevant information improves performance with lifts in precision by up to 20x. 

We analyzed the impact of differing document lengths in the prompt. Recall increased by~10\%  when going from three chunks to five. Including the entire document boosted precision to 99\% and recall to 84\%. 
While long context windows (a feature of some state-of-the-art llms) seem to perform significantly better we argue that large, complex and diverse collections of documents may still benefit from chunking and summarization to strike a balance between performances and running costs.

\begin{table}
\centering
\caption{Performance for generating Failure Locations on test dataset n = 12 (SSEE similarity threshold 0.8 for all experiments) }
\begin{tabular}{cccc}
\hline
       & Context & SSEE     & SSEE  \\
Method & Length  & Precision & Recall \\ 
\hline
Zero-shot   & --  &  0.05       & 0.11  \\
RAG system      & 3 chunks   &  0.66      & 0.62\\
RAG system      & 5 chunks    &  0.66      & 0.72\\
RAG system     & Long          &  0.99      & 0.84\\
\hline
\end{tabular}
\label{tab:fl_results}
\end{table}

\section{Conclusions and Future Work}

This paper demonstrated an interactive LLM-based application for creating
new FMEAs from reference documents.
As future work, we plan to explore approaches that draw context from both reference documents and existing FMEA databases.
Multi-modal foundation models also show great promise for accessing information in equipment drawings and process flow diagrams.

\bibliographystyle{splncs04}
\bibliography{main}

\begin{thebibliography}{10}
\providecommand{\url}[1]{\texttt{#1}}
\providecommand{\urlprefix}{URL }
\providecommand{\doi}[1]{https://doi.org/#1}

\bibitem{auer2024doclingtechnicalreport}
Auer, C., Lysak, M., Nassar, A., Dolfi, M., Livathinos, N., Vagenas, P., Ramis,
  C.B., Omenetti, M., Lindlbauer, F., Dinkla, K., Mishra, L., Kim, Y., Gupta,
  S., de~Lima, R.T., Weber, V., Morin, L., Meijer, I., Kuropiatnyk, V., Staar,
  P.W.J.: Docling technical report (2024),
  \url{https://arxiv.org/abs/2408.09869}

\bibitem{cooper2015FMEAs}
Cooper, H.C.: Capture all critical failure modes into fmea in half the time
  with a simple decomposition table (actual case study savings = 4,206,000).
  In: RAMS. pp.~1--6 (2015)

\bibitem{elHassani2024}
El~Hassani, I., Masrour, T., Kourouma, N., Motte, D., Tav{\v{c}}ar, J.:
  Integrating large language models for improved failure mode and effects
  analysis (fmea): a framework and case study. Proceedings of the Design
  Society  \textbf{4},  2019--2028 (2024)

\bibitem{granite2024ver3}
Granite~Team, I.: Granite 3.1 language models.
  \url{https://github.com/ibm-granite/granite-3.1-language-models/} (December
  2024)

\bibitem{Hodkiewicz2021}
Hodkiewicz, M., Klüwer, J.W., Woods, C., Smoker, T., Low, E.: An ontology for
  reasoning over engineering textual data stored in fmea spreadsheet tables.
  Computers in Industry  \textbf{131},  103496 (2021).
  \doi{https://doi.org/10.1016/j.compind.2021.103496},
  \url{https://www.sciencedirect.com/science/article/pii/S0166361521001032}

\bibitem{lewis2021rag}
Lewis, P., Perez, E., Piktus, A., Petroni, F., Karpukhin, V., Goyal, N.,
  Küttler, H., Lewis, M., tau Yih, W., Rocktäschel, T., Riedel, S., Kiela,
  D.: Retrieval-augmented generation for knowledge-intensive nlp tasks (2021),
  \url{https://arxiv.org/abs/2005.11401}

\bibitem{fmea_builder_ijcai_2024}
Lynch, K., Lorenzi, F., Sheehan, J., Kabakci-Zorlu, D., Eck, B.: Fmea builder:
  Expert guided text generation for equipment maintenance. In: AI for Critical
  Infrastructure Workshop at 33rd International Joint Conference on Artificial
  Intelligence {(IJCAI-24)} (2024)

\bibitem{Lynch2025}
Lynch, K., Lorenzi, F., Sheehan, J.D., Kabakci-Zorlu, D., Eck, B.: Structured
  document generation for industrial equipment. Proceedings of the AAAI
  Conference on Artificial Intelligence  \textbf{39}(28),  28850--28856 (Apr
  2025). \doi{10.1609/aaai.v39i28.35150},
  \url{https://ojs.aaai.org/index.php/AAAI/article/view/35150}

\bibitem{srt2004}
Rausand, M., Høyland, A.: System Reliability Theory: Models, Statistical
  Methods, and Applications (2nd ed.). Wiley (2004)

\bibitem{sharma:JARASS-2018}
Sharma, K.D., Srivastava, S.: Failure mode and effect analysis (fmea)
  implementation: a literature review. J Adv Res Aeronaut Space Sci
  \textbf{5}(1-2),  1--17 (2018)

\bibitem{Xia2024}
Xia, Y., Jazdi, N., Weyrich, M.: Enhance fmea with large language models for
  assisted risk management in technical processes and products. pp.~1--4 (09
  2024). \doi{10.1109/ETFA61755.2024.10710996}

\end{thebibliography}

\end{document}